\documentclass[pra,twocolumn,superscriptaddress,floatfix]{revtex4}

\usepackage{graphics}           
\usepackage{bm}                 
\usepackage{amsmath}            
\usepackage{amsfonts}
\usepackage{amssymb}
\usepackage{latexsym}           

\newcommand{\ket}[1]{\mbox{$|#1\rangle$}}
\newcommand{\bra}[1]{\mbox{$\langle#1|$}}
\newcommand{\braket}[2]{\mbox{$\langle#1|#2\rangle$}}

\begin{document}

\title{Controlling discrete quantum walks: coins and intitial states}

\author{Ben Tregenna}
\author{Will Flanagan}
\author{Rik Maile}
\affiliation{Optics Section, Blackett Laboratory, Imperial College,
London, SW7 2BW, United Kingdom.}
\author{Viv Kendon}
\email{Viv.Kendon@ic.ac.uk}
\affiliation{Optics Section, Blackett Laboratory, Imperial College,
London, SW7 2BW, United Kingdom.}
\affiliation{Mathematical Sciences Research Institute, 1000 Centennial Drive,
Berkeley, CA 94720-5070, USA}

\date{May 13, 2003}

\begin{abstract}
In discrete time, coined quantum walks, the coin degrees of freedom offer
the potential for a wider range of controls over the evolution of
the walk than are available in the continuous time quantum walk.
This paper explores some of the possibilities on regular graphs,
and also reports periodic behaviour on small cyclic graphs.
\end{abstract}



\maketitle


\section{Introduction}
\label{sec:intro}

Quantum walks are analogs of classical random walks,
designed primarily with the aim of finding quantum algorithms that
are faster than classical algorithms for the same problem.
There are two distinct types of quantum walks, corresponding to
classical random walks with discrete or continuous time (but both
taking place in a discrete space).  Continuous time quantum walks were first 
introduced in 1997 by Farhi and Gutmann \cite{farhi98a}.
Discrete time quantum walks with a quantum coin appeared in the early 1990s
in work by Y Aharonov \textit{et al.}~\cite{aharonov92a},
then were developed as quantum cellular automata by Meyer
\cite{meyer96a,meyer96b,meyer96c} in 1996.
The first explicitly algorithmic context for coined quantum walks came from
D Aharonov \textit{et al.}~\cite{aharonov00a} and
Ambainis \textit{et al.}~\cite{ambainis01a} in 2000.

Two algorithms for quantum walks
have recently been presented.  Childs \textit{et al.}~\cite{childs02a}
prove that a continuous time quantum walk can find its way
across a special type of graph exponentially faster than any
classical algorithm, and Shenvi \textit{et al.}~\cite{shenvi02a}
prove that a discrete time, coined quantum walk can equal Grover's
search algorithm, by finding a marked item in an unsorted database
with a quadratic speed up over the best known classical algorithm.
These results are extremely promising, but still a long way from
the diversity of problems that classical random walks provide
the best known solutions for, such as approximating the permanent of
a matrix \cite{jerrum01a}, finding satisfying assignments to Boolean
expressions ($k$SAT with $k>2$) \cite{schoning99a},
estimating the volume of a convex body \cite{dyer91a},
and graph connectivity \cite{motwani95}.
Classical random walks underpin many standard methods in computational
physics, such as Monte Carlo simulations, so a more efficient quantum
alternative would presumably widen the potential application of
quantum computers to problems in physics.

In much the same way as we now know almost everything about the
properties and possible states of two qubits, though quantum computers
will clearly need far more than two qubits to be useful, the simple
quantum walk on a line has now been well studied, see for example
Refs.~\cite{ambainis01a,bach02a,yamasaki02a,kendon02c,brun02b,brun02c,carteret03a},
though there is no suggestion that it will lead to useful quantum walk
algorithms by itself.
The quantum walk on a cycle is a step closer to algorithms.
The $N$-cycle is the Cayley graph of the cyclic group of size $N$, and 
in addition to proving that the coined quantum walk on a cycle
has a time-averaged mixing time almost quadratically faster than a
classical random walk, D Aharonov \textit{et al.}~\cite{aharonov00a}
also provided a lower bound on the time-averaged mixing times for
quantum walks on general graphs of bounded degree,
suggesting a quadratic improvement
over classical random walks is the best that can be achieved.
Moore and Russell \cite{moore01a} solved both discrete and continuous
time quantum walks on the hypercube of size $N$, showing that
both have an instantaneous mixing time linear in $N$, logarithmically
faster than classical random walks.  However, they also
showed that time-averaged mixing times on the hypercube are slower
than classical, the continuous time walk never mixes in the sense
of the time-averaged definition.
Kempe \cite{kempe02a} proved that a quantum walk can travel from
one corner of a hypercube to the opposite corner exponentially
faster than a classical random walk, however,
there are other classical algorithms that can do this task efficiently
so this does not provide a quantum advantage over classical.
For a recent survey of quantum walks and a more complete
list of references, see Kempe \cite{kempe03a}.

So far, though the published literature on discrete and continuous time
quantum walks tends to treat different problems, the evidence suggests
that both can accomplish the same tasks.  They are clearly not
exactly equivalent, and the computational equivalence observed
depends on choosing an
appropriate form for the coin operator for the discrete time walks.
This raises the possibility that different choices of coin operator
could perform other useful tasks that aren't easily accessible within
the continuous time quantum walk model.
In this paper, we present a study of the properties of different coin
operators using both analytical and numerical methods.
Our results are of interest both in themselves as examples of quantum
dynamics, and as potential ingredients for quantum algorithms.
The paper is organized as follows.  After setting up our notation,
we discuss the possibilities for graphs of degree two, three, and four 
in Secs.~\ref{sec:2d}, \ref{sec:3d}, and \ref{sec:4d} respectively.
In Sec.~\ref{sec:hd} we briefely mention graphs of higher degree and in
Sec.~\ref{sec:pd} we describe periodic quantum walks on cyclic graphs.

\subsection{Notation for a general quantum walk}

A general coined quantum walk on a $d$-regular graph
needs a coin Hilbert space, $\mathcal{H}_d$ with $d$ the degree of
each vertex in the graph on which the walk takes place,
and a position Hilbert space $\mathcal{H}_N$ with $N$ the
number of vertices in the graph (which can be infinite).
The dynamics of the walk are controlled by a
coin flip operator $\mathbf{C}$ that acts on the coin Hilbert space,
and a conditional shift operator $\mathbf{S}$
that shifts the particle position according to the state of the coin.
Together, $\mathbf{U}\equiv\mathbf{S}(\mathbf{C}\otimes\mathbf{I}_N)$
is the unitary operator for one step of the walk.
If the particle and coin start in state $\ket{\psi_0}$, the state
of the system after $t$ steps of the walk is
$\ket{\psi_t} = \mathbf{U}^t\ket{\psi_0}$.

A powerful technique for the solution of classical random walks that
generalises well to the quantum case is that of Fourier
transformation.  When the walk occurs on the Cayley graph of some
group, the quantum walk simplifies greatly on consideration of the
Fourier space of the particle \cite{aharonov00a,ambainis01a}. 
Quantum walks on the infinite line, $N$-cycle and the
hypercube admit this type of solution. 
An alternative method using path counting (path integrals) was also
presented in \cite{ambainis01a} and further refined in \cite{carteret03a}.

\section{Graphs of Degree Two}
\label{sec:2d}

We will consider the simplest examples first, 
coined quantum walks on the line and the cycle.
The walk on the line has already been analysed in detail
and the equivalence of all unbiased coin operators noted by
several authors \cite{ambainis01a,bach02a,yamasaki02a}.
We first review these calculations, since the notation
and results will be used in our analysis of the walk on the $N$-cycle.

\subsection{Quantum walk on an infinite line}
\label{ssec:2dline}

The most general two dimensional unitary coin operator
$\mathbf{C}_2^{(\text{gen})}$ can be written as a $2\times 2$ matrix
\begin{equation}
\mathbf{C}_2^{(\text{gen})}=\left( \begin{array}{cc}
        \sqrt{\rho}& \sqrt{1-\rho}e^{i\theta}\\
        \sqrt{1-\rho}e^{i\phi}& -\sqrt{\rho}e^{i(\theta+\phi)}
        \end{array} \right),
\label{eq:genH}
\end{equation}
where $0\le\theta,\phi\le\pi$ are arbitrary angles, $0\le\rho\le 1$,
and we have removed an irrelevant global phase so as to leave the
leading diagonal element real.  The Hadamard coin operator is
obtained with $\rho=1/2$ and $\theta=\phi=0$.
The parameter $\rho$ thus controls the bias of the coin, $\rho=1/2$
being a fair coin that chooses each of the two possible directions
$\ket{R}$ (right) and $\ket{L}$ (left) with equal probability.
Trivial cases $\rho=0$,1 give oscillatory motion and uniform motion 
respectively.
The Fourier transformation is performed only over the particle Hilbert space,
\begin{equation}
\ket{\tilde{\psi}(k,t)} = \sum_x \ket{\psi(x,t)}e^{ikx}.
\label{eq:ftline}
\end{equation}
Here the state vectors \ket{\psi(x,t)} and \ket{\tilde{\psi}(k,t)} are
two component vectors, with the first component being the amplitude of the
right moving part and the second component being that of the left moving
part, with $k\in[0,2\pi)$.
Using the general form of the coin transition matrix for a one
dimensional walk, Eq.~(\ref{eq:genH}), a single step of the walk
becomes
\begin{equation}
\ket{\tilde{\psi}(k,t+1)} = \mathbf{C}_k^{(\text{gen})}\ket{\tilde{\psi}(k,t)},
\label{eq:ftevolve}
\end{equation}
where $\mathbf{C}_k^{(\text{gen})}$ is a $2\times 2$ matrix acting on the coin Hilbert
space,
\begin{equation}
\mathbf{C}_k^{(\text{gen})}=\left( \begin{array}{cc}
        \sqrt{\rho}e^{ik}& \sqrt{1-\rho}e^{i(k+\theta)}\\
        \sqrt{1-\rho}e^{i(-k+\phi)}& -\sqrt{\rho}e^{i(-k+\theta+\phi)}
        \end{array} \right).
\end{equation}
This matrix may be diagonalised, yielding eigenvalues
\begin{equation}
\lambda_k^\pm=\pm e^{i\delta}e^{\pm i\omega_k},
\end{equation}
where $\delta=(\theta+\phi)/2$ and
\begin{equation}
\sin(\omega_k)=\sqrt{\rho}\sin(k-\delta).
\label{eq:omegakline}
\end{equation}
The associated eigenvectors are
\begin{equation}
\ket{\tilde{\xi}_k^\pm}=\frac{1}{n_k^{\pm}}\left( \begin{array}{c}
        e^{ik}\\
	e^{-i\theta}(\lambda_k^\pm - \sqrt{\rho}e^{ik})/\sqrt{1-\rho}\\
        \end{array} \right),
\end{equation}
with the normalisation factor $n_k^{\pm}$ given by
\begin{equation}
(n_k^{\pm})^2=2\left\{1\mp\sqrt{\rho}\cos(k-\delta\mp\omega_k)\right\}/(1-\rho)
\end{equation}
For a general unbiased initial coin state,
$\ket{\psi(x,0)}=\sqrt{\eta}(\ket{R}+e^{i\alpha}\sqrt{1-\eta}\ket{L})\otimes\ket{0}$,
the Fourier components at $t=0$ can be found from Eq.~(\ref{eq:ftline}),
\begin{equation}
\ket{\tilde{\psi}(k,0)}=\left( \begin{array}{c}
        \sqrt{\eta}\\e^{i\alpha}\sqrt{1-\eta}
        \end{array} \right)\otimes\ket{k} ~~~~\forall k.
\end{equation}
Collecting all these pieces together, it is possible to write down the
Fourier components at all later times $t$,
\begin{equation}
\ket{\tilde{\psi}(k,t)} = (\mathbf{C}_k^{(\text{gen})})^t\ket{\tilde{\psi}(k,0}.
\end{equation}
Expressing $\mathbf{C}_k^{(\text{gen})}$ in terms of its eigenvalues and eigenvectors,
$(\mathbf{C}_k^{(\text{gen})})^t= (\lambda_k^+)^t\ket{\tilde{\xi}_k^+}\bra{\tilde{\xi}_k^+}
+ (\lambda_k^-)^t\ket{\tilde{\xi}_k^-}\bra{\tilde{\xi}_k^-}$, gives
\begin{eqnarray}
\ket{\tilde{\psi}(k,t)}
&=& (\lambda_k^+)^t\ket{\tilde{\xi}_k^+}\braket{\tilde{\xi}_k^+}{\tilde{\psi}(k,0)}\nonumber\\
&+& (\lambda_k^-)^t\ket{\tilde{\xi}_k^-}\braket{\tilde{\xi}_k^-}{\tilde{\psi}(k,0)}
\label{eq:linesol}
\end{eqnarray}
The coefficients of $\ket{\tilde{\xi}_k^{\pm}}$ are given by
\begin{eqnarray}
(\lambda_k^{\pm})^t\!\!\!\!&&\braket{\tilde{\xi}_k^{\pm}}{\tilde{\psi}(k,0)}
=\frac{(\lambda_k^{\pm})^t}{n_k^{\pm}}e^{-ik}\left\{
\text{\rule[-1em]{0em}{2em}}\sqrt{\eta}\right.\nonumber\\
	&&\left.-\sqrt{\frac{1-\eta}{1-\rho}} e^{i(\theta+\alpha)}
        (\sqrt{\rho} \mp e^{i(k-\delta)}e^{\mp i\omega_k})\right\}.
\label{eq:linegfc}
\end{eqnarray}
All the subsequent statistics for the probability distribution may be
found by inverting the Fourier transform and applying standard methods
from complex analysis \cite{ambainis01a,carteret03a}.
However, the question of the effect of the extra
degrees of freedom $\alpha$ and $\eta$ pertaining to the quantum coin,
and $\phi$, $\theta$ and $\rho$ in the coin operator,
may be answered directly from Eq.~(\ref{eq:linegfc}).
The parameters $\eta$ and $\rho$ appear in a non-trivial way and thus affect
the subsequent evolution of the walk, but the phase factor
$\alpha$ occurs solely with the phase $\theta$ in the coin flip
operator as the combination $(\theta+\alpha)$.
The other influences on the evolution from the phases in the coin flip matrix
come from the factor $e^{i\delta t}$ in the eigenvalues, which is a global
phase and therefore doesn't affect observable quantities, and
from phases of $e^{i(k-\delta)}$ (explicitly and in $\omega_k$),
which disappear when $k$ is integrated over its full range during
the inverse Fourier transform.
Thus, for any given $\theta$ in the coin operator, one may choose
an $\alpha$ so as to give the full range of possible evolutions.
This has been noted by several authors, \cite{ambainis01a,bach02a,yamasaki02a}.
For the walk on a line, without loss of generality,
one may thus restrict the coin operator to one with
real coefficients, and obtain the full range of behaviour by
choosing different initial coin states.  Further restricting to unbiased coins
($\rho=1/2$), the Hadamard coin
\begin{equation}
\mathbf{C}_2^{(\text{H})}=\left( \begin{array}{rr}
        1 & 1 \\
        1 & -1 \\
        \end{array} \right),
\label{eq:had}
\end{equation}
is thus the only possible type of coin for the quantum walk on a line.

\begin{figure}
    \begin{minipage}{\columnwidth}
            \begin{center}
            \resizebox{\columnwidth}{!}{\includegraphics{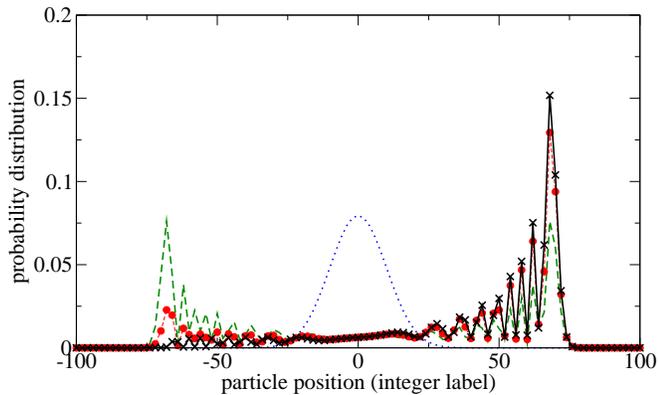}}
            \end{center}
            \caption{Asymmetric distributions obtained with various
		initial states and Hadamard coin for the walk on a
		line after 100 steps:
		coin bias $\eta \simeq 0.85$ (crosses),
		coin state $\ket{R}$ (circles),
		with symmetric (dashed) and classical (dotted) for comparison.
		Only even-numbered positions are plotted,
		since the distribution is zero on odd-numbered positions.}
            \label{fig:l100-skew}
    \end{minipage}
\end{figure}
The asymmetry of the distribution obtained for an initial coin state
of $\ket{R}$ or $\ket{L}$ is now well-known, and is also
obtained for unbiased initial states with $\alpha = 0$ or $\pi$.
However, it is possible to create an even more biased distribution
using the Hadamard (unbiased) coin operator,
by choosing a biased initial state with $\eta \simeq 0.85$, i.e.,
$\ket{\psi_0} = (\sqrt{0.85}\ket{R}+\sqrt{0.15}\ket{L})\otimes\ket{0}$.
This is shown in Fig.~\ref{fig:l100-skew}, along with the distributions
for $\ket{R}$, and symmetric quantum and classical distributions for
comparison.
The asymmetry of the distribution can be quantified by examining
the third moment, which we take about the origin, i.e., with reference
to the initial location of the particle, normalised by
the second moment, $\langle x^3\rangle / \langle x^2 \rangle^{3/2}$.
This quantity is just greater than one for the initial state with
$\eta \simeq 0.85$, and around 0.7 for an initial state of $\ket{R}$
(obtained numerically, for analytic formulae see Konno \cite{konno02b,konno03a},
the value of $\eta$ comes from $\cos(\pi/8)\simeq0.85$).
By simply changing the phase by $\pi$ to
$\sqrt{0.85}\ket{R}-\sqrt{0.15}\ket{L}$, the distribution becomes symmetric.
Comparing this with the unbiased coin initial state
$(\ket{R} + i\ket{L})/\sqrt{2}$ that also gives a symmetric evolution,
and noting that the Hadamard operator is real, so any component with phase
$i$ remains orthogonal to any real component, we can see that there are two
distinct ways of arriving at a symmetric quantum walk on a line.
Initial states $\sqrt{0.85}\ket{R}-\sqrt{0.15}\ket{L}$ (biased)
and $(\ket{R} + i\ket{L})/\sqrt{2}$ (symmetric) give almost identical
probability distributions, but the former is obtained by interference and
the latter by combining probabilities from two mirror image orthogonal
components.

This gives us our first insights into how to use the coin to control the 
walk.  The quadratic speed up in the spreading of the quantum walk over
classical is unaffected by the choice of initial state or coin operator:
the speed up comes solely from the coherent wave motion along the line.
This is made clearest by noting that a maximally mixed initial coin state
also produces a symmetric distribution like the previous two examples.
We can then control whether (and to what degree)
the waves will interfer constructively, destructively, or not at all,
by choosing the phase and bias of the coin initial state.

\subsection{Quantum Walk on a $N$-Cycle}
\label{ssec:2dcyc}

The walk on a $N$-cycle is the same as the walk on the line but
with the particle position taken as $x$ (mod $N$).
It is also amenable to solution in the Fourier
basis. The finite state space of the particle gives rises to a
discrete, finite momentum space defined by
\begin{equation}
\ket{\tilde{\psi}_N(k,t)} = \frac{1}{\sqrt{N}}\sum_{x=0}^N \ket{\psi_N(x,t)}e^{2\pi 
ikx/N},
\end{equation}
for $k\in\{0,1,...N-1\}$.
From here it is possible to proceed in a similar manner to that shown in
Sec.~\ref{ssec:2dline} for a walk on an infinite line.
Equation (\ref{eq:ftevolve}) may be used again, with the discrete
version of $\mathbf{C}_k^{(\text{gen})}$ given by
\begin{equation}
\mathbf{C}_k^{(N)}=\left( \begin{array}{cc}
        \sqrt{\rho}e^{2\pi ik/N}& \sqrt{1-\rho}e^{i(2\pi k/N+\theta)}\\
        \sqrt{1-\rho}e^{i(-2\pi k/N+\phi)}&
				 -\sqrt{\rho}e^{i(-2\pi k/N+\theta+\phi)}
        \end{array} \right).
\end{equation}
This may again be diagonalised, yielding eigenvalues
\begin{equation}
\lambda_k^\pm=\pm e^{i\delta}e^{\pm i\omega_k^{(N)}},
\label{eq:lambdakcycle}
\end{equation}
where now
\begin{equation}
\sin(\omega_k^{(N)})=\sqrt{\rho}\sin(2\pi k/N-\delta),
\label{eq:omegakcycle}
\end{equation}
compare Eq.~(\ref{eq:omegakline}).
The possible solutions for $\omega_k^{(N)}$ are bounded by
$\sin^{-1}(\sqrt{\rho})$, e.g., for $\rho=1/2$,
there are two solutions for $\omega_k^{(N)}$
one in each of the regions $[\pi/4,3\pi/4]$ and $[-\pi/4,-3\pi/4]$.
The first solution corresponds to $\lambda^+_k$ and the second to
$\lambda^-_k$.

Classically, a random walk on a cycle tends to a uniform distribution over
all points on the cycle at long times.
Since the quantum walk is unitary and reversible, it
never reaches a uniform distribution, the initial state influences the
particle's dynamics at all later times.
However, we can define a time-averaged distribution \cite{aharonov00a}
which does tend to a limiting value for large $T$,
\begin{equation}
\overline{P}(x,T)=\frac{1}{T}\sum_{t=0}^{T-1}P(x,t),
\label{eq:limdef}
\end{equation}
where $P(x,t)= |U^t\ket{\psi_N(x,0)}|^2$.
It is proven in \cite{aharonov00a} that
\begin{eqnarray}
\lim_{T\to\infty}\overline{P}(x,T)&=&\sum_{v,u\in \lambda_v=\lambda_u}
\braket{\psi_N(x,0)}{\phi_v^\pm}\braket{\phi_u^\pm}{\psi_N(x,0)}\nonumber\\
&\times&\sum_a\braket{x,a}{\phi_v^\pm}\braket{\phi_u^\pm}{x,a},
\label{eq:limproof}
\end{eqnarray}
where the sum is taken only over degenerate eigenvectors of the
position space evolution matrix $\mathbf{U}$, which are denoted by
\ket{\phi_v^\pm}, \ket{\phi_u^\pm}, with eigenvalues
$\lambda_v^\pm=\lambda_u^\pm$.
The general initial state is once again
$\ket{\psi_N(x,0)}=\sqrt{\eta}(\ket{R}
+e^{i\alpha}\sqrt{1-\eta}\ket{L})\otimes\ket{0}$. 
For the walk on a $N$-cycle,
the eigenvectors of $\mathbf{U}$ in the position basis are given by
$\ket{\phi_v^\pm}=\ket{\xi_k^\pm}\otimes\ket{\chi_k}$,
where the \ket{\xi_k^\pm} are the eigenvectors of the matrix $\mathbf{C}_k^{(N)}$
and $\ket{\chi_k}=\tfrac{1}{\sqrt{N}}\sum_x e^{2\pi i kx/N}\ket{x}$,
i.e., a discrete Fourier transform of the usual particle position basis
states.
(We omit labels of $N$ from $\ket{\xi_k^\pm}$ and $\ket{\chi_k}$ to keep the notation less cluttered.)
The associated eigenvalues of $\mathbf{U}$ are (by
construction) equal to those of the matrices $\mathbf{C}_k^{(N)}$, namely
$\lambda_k^\pm$. Hence we can rewrite Eq.~(\ref{eq:limproof}) in terms of
the eigenvectors of $\mathbf{C}_k^{(N)}$,
\begin{eqnarray}
\lim_{T\to\infty}\overline{P}(x,T) &=& \sum_{a,k,j,b,c}
\braket{\psi_N(x,0)}{\chi_k,\xi_k^b}\braket{\chi_j,\xi_j^c}{\psi_N(x,0)}\nonumber\\
&\times&\braket{x,a}{\chi_k,\xi_k^b}\braket{\chi_j,\xi_j^c}{x,a}.
\label{eq:limk}
\end{eqnarray}
The sum is taken over $k,j,b$ and $c$ such that $\lambda_k^b=\lambda_j^c$.
It was shown in \cite{aharonov00a} that since the \ket{\chi_k} induce a
uniform distribution over the nodes, the limiting distribution will
also be uniform if all eigenvalues are distinct. The eigenvalues are
degenerate in the general case if there exist non-trivial solutions for
\begin{equation}
\sin(2\pi k/N-\delta) =\sin(2\pi j/N -\delta).
\end{equation}
This equation has solutions $k=j$ and $k+j \text{(mod $N$)} =N/\pi(\delta+\pi/2)$. The first
is trivial, but whether the second solution has roots depends on
the coin flip operator and on $N$. 
For example, when a Hadamard coin is used, $\theta=\phi=\delta=0$ so the
condition becomes $k+j \text{(mod $N$)} =N/2$ which has roots only for even $N$.
Thus for a Hadamard walk, cycles with an odd number of
nodes converge to the uniform distribution and those with an even
number converge to a non-uniform distribution derived below.
However, for a given $N$, the 
coin flip operator with $(\delta+\pi/2)=\pi/N$ gives roots when
$k+j \text{(mod $N$)}=1$ which
always has a solution, leading to a non-uniform limiting distribution.
Conversely, if $\delta$ is not a rational 
multiple of $\pi$, there can be no solutions, and so the walk will
always mix to the uniform distribution. Thus, by appropriate choice of
coin operator, a walk on any size cycle can be made to converge either
to a uniform or to a non-uniform probability distribution.
This is in direct contrast to the classical case, in which the properties
of the limiting distribution depend solely on the form of the graph.

We note that the limit as the cycle size $N\rightarrow\infty$ leads to
the condition $\delta =-\pi/2$ for a non-uniform limiting distribution.
This gives $\theta+\phi = -\pi$, the simplest unbaised coin operator
corresponding to this is
\begin{equation}
\mathbf{C}_2^{(\text{nu})}=\frac{1}{\sqrt{2}}\left( \begin{array}{rr}
        1 & -i \\
        -i & 1
        \end{array} \right).
\end{equation}
However, the practical limit of infinite cycle size is the walk on a line,
where the the opposite edges of the walk never meet, and conditions
for non-uniform distributions are not meaningful.

It is possible to derive the limiting distribution when there exist
degenerate eigenvalues of the evolution operator $\mathbf{U}$.
In these cases, the summation in Eq.~(\ref{eq:limk}) contains
two distinct types of terms, those for
which $k=j$ and those for which $j=N/\pi(\delta+\pi/2)-k\equiv\Phi-k$,
\begin{eqnarray}
\lim_{T\to\infty}\!\!\!\!&&\!\!\!\!\overline{P}(x,T) = \sum_{a,k,b}
{\Big [}|\braket{\psi_N(x,0)}{\chi_k,\xi_k^b}|^2
 |\braket{x,a}{\chi_k,\xi_k^b}|^2\nonumber\\
&+&\braket{\psi_N(x,0)}{\chi_k,\xi_k^b}\braket{\chi_{\Phi-k},\xi_{\Phi-k}^{-b}}{\psi_N(x,0)}\nonumber\\
&&\times\braket{x,a}{\chi_k,\xi_k^b}\braket{\chi_{\Phi-k},\xi_{\Phi-k}^{-b}}{x,a}{\Big ]}.
\end{eqnarray}
Using $|\braket{x}{\chi_k}|^2 = 1/N$, $\sum_a|\braket{a}{\xi_k^b}|^2 = 1$
and $\sum_{k,b}|\braket{\psi_N(x,0)}{\chi_k,\xi_k^b}|^2=1$,
the first term is easily seen to be the uniform distribution ($1/N$).
In the second term, the factor that determines the
form of the limiting distribution is 
\begin{equation}
\braket{x}{\chi_k}\braket{\chi_{\Phi-k}}{x}
= e^{4\pi i x/N(k - N\delta/2\pi -N/4)},
\end{equation}
which controls the sign of the terms in the sums.
When $x=0$, all the terms in the summation are positive, leading to a
spike in the distribution about the origin.
Similarly, if $x=N/2$, the phase of each term is 
$2\pi(k - N\delta/2\pi - N/4)$ so the terms add coherently
(remember $(N\delta/\pi - N/2)$ is an integer).
Specifically, for the Hadamard coin, $\delta=0$
and the contribution to the sum is positive if $N/2$ is even,
i.e., $N$ is divisible by four,
or negative if $N/2$ is odd, leading to a minimum.
The Hadamard case has been independently calculated in more detail
by Bednarska \textit{et al.}~\cite{bednarska03a}, who also explore some
possibilities for highly non-uniform limiting distributions generated by
initial states superposing several particle positions.

The effects of different coin flip operators have received little attention
in the literature to date, perhaps due to the minimal effect they have
for a walk on a line.
However, for quantum walks containing closed cycles,
the choice of coin flip operator determines which phase the wavefronts have
when they meet up with each other, selecting between whether constructive or
destructive interference occurs.
Note that in \cite{kendon02c} it was shown that decoherence
in a walk on a cycle causes all initial states and coin operators
to mix to the uniform distribution even while there is still a
clear quantum speed up over the classical mixing times.
The coherence required for non-uniform limiting distributions is thus
much more stringent than that required for a quantum speed up of
the mixing time over classical.  This suggests that in order to use
the effects non-uniform limiting distributions it
will be more useful if they have properties that
can be measured after relatively few steps of the walk,
rather than waiting for long times.

\section{Graphs of Degree Three}
\label{sec:3d}

\begin{figure}
    \begin{minipage}{\columnwidth}
	    \begin{raggedright}
            \resizebox{0.95\columnwidth}{!}{\includegraphics{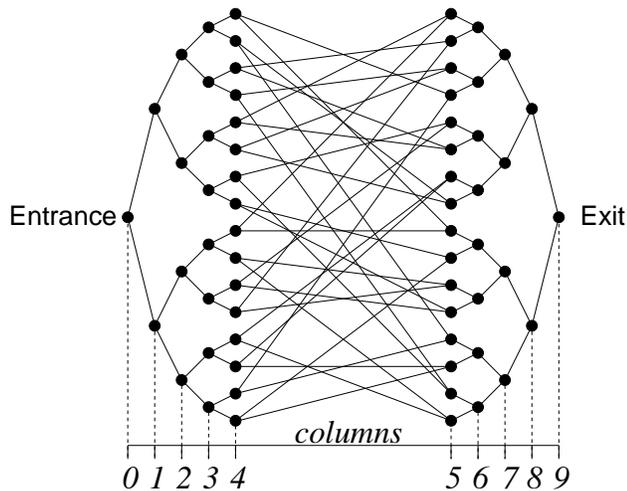}}
	    \end{raggedright}
	    \hfill
            \caption{``Glued trees'' graph used in the algorithm of
		Ref.~\cite{childs02a}.  Example shown is for $N=4$, with
		$2N+1=9$ columns labeled at the bottom of the figure,
		and $2(2^{(N+1)}-1)=62$ nodes.
		The task is to travel from entrance
		to exit without getting lost in the randomly joined middle
		section of the graph.  The gap between columns 4 and 5 is for
		clarity in the figure and is not significant in the algorithm.}
            \label{fig:tree4}
    \end{minipage}
\end{figure}

\begin{figure}
    \begin{minipage}{\columnwidth}
            \begin{center}
            \resizebox{\columnwidth}{!}{\includegraphics{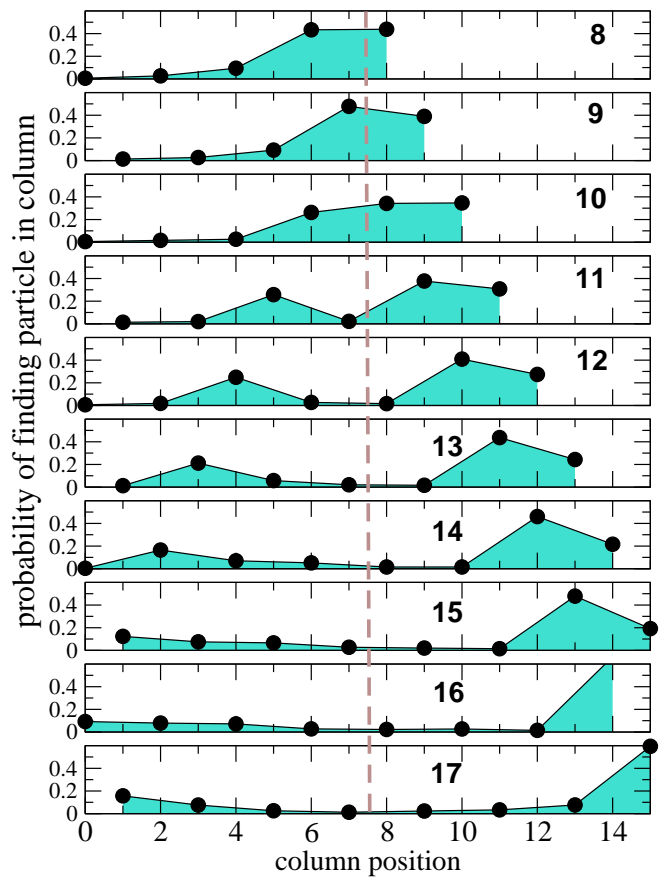}}
            \end{center}
            \caption{Distribution over columns of the ``glued trees'' graph
		of Ref.~\cite{childs02a} with a discrete time walk using
		a Grover coin.  This is for a graph of size $N=7$
		(with $2N+1=15$ columns).
		The vertical dashed line indicates the position
		of the random join between the two trees.
		The quantum walk reaches the
		far end in just 17 steps, with probability around 0.6
		(same as the continuous time version).}
            \label{fig:7treegrov2}
    \end{minipage}
\end{figure}

Regular lattices of degree three have been studied briefly numerically
\cite{mackay01a}, where the spreading rate was shown to be faster than
classical.
The ``glued trees'' graph used for the algorithm presented in
Ref.~\cite{childs02a} is also of degree three apart from the special
start and end points that form the roots of the two binary
trees, see Fig.~\ref{fig:tree4}.
This structure is highly symmetric, despite the random connections in the
middle, and provided a symmetric initial state is used at the entrance node,
the whole quantum walk process can be mapped to a walk on a line (the column
positions shown in Fig.~\ref{fig:tree4}) with different biases in the probabilities for moving right or left at each step.
Childs \textit{et al.}~\cite{childs02a} use a continuous time walk for their
algorithm, but if a three dimensional coin based on Grover's diffusion operator
with elements $2/d-\delta_{ij}$,
\begin{equation}
\mathbf{C}_3^{(G)}=\frac{1}{3}\left( \begin{array}{rrr}
        -1 & 2 & 2\\
	 2 &-1 & 2\\
	 2 & 2 &-1\\
        \end{array} \right),
\label{eq:grov3}
\end{equation}
is used with a discrete walk, the amplitude also
interfers constructively in the right way to reach the opposite
root of the trees quickly with high probability \cite{watrous02a,kendon02c},
see Fig.~\ref{fig:7treegrov2}.

\begin{figure}
    \begin{minipage}{\columnwidth}
            \begin{center}
            \resizebox{\columnwidth}{!}{\includegraphics{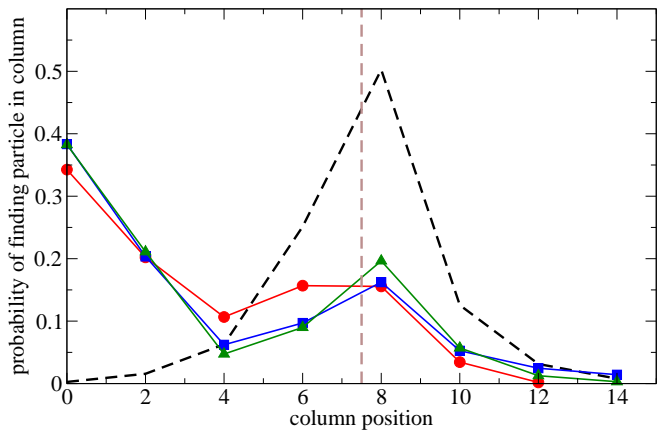}}
            \end{center}
            \caption{As Fig.~\ref{fig:7treegrov2} but with DFT coin,
		after 12 (circles), 60 (squares), 120 (triangles) steps
		of the walk.  A classical random walk after 120 steps
		is shown dashed.}
            \label{fig:7treedft}
    \end{minipage}
\end{figure}
The Grover coin is biased but symmetric.
The DFT (discrete Fourier transform) coin is unbiased,
but asymmetric in that you cannot interchange the
labels on the directions without changing the coin operator.
For $d=3$, it looks like
\begin{equation}
\mathbf{C}_3^{(D)}=\frac{1}{\sqrt{3}}\left( \begin{array}{ccc}
        1 & 1 & 1\\
	1 & e^{i\omega} & e^{-i\omega}\\
	1 & e^{-i\omega} & e^{i\omega}\\
        \end{array} \right),
\label{eq:DFT3}
\end{equation}
where $e^{i\omega}$ and $e^{-i\omega}$ are the complex cube roots of unity,
For $d=2$, the DFT coin reduces to the Hadamard coin, Eq.~(\ref{eq:had}).
If a DFT coin is used instead of a Grover coin on the ``glued trees''
graph, this keeps the amplitude near the starting point 
and the walk does not spread out even as far as a classical random walk,
see Fig.~\ref{fig:7treedft}.
While this is not useful in the context of the ``glued trees'' problem,
it is still highly non-classical behaviour, and with the right problem and
initial coin state, the DFT coin operator may find its place in a
useful quantum walk algorithm.


\section{Graphs of Degree Four}
\label{sec:4d}

Quantum walks on regular two dimensional lattices have been
investigated numerically by Mackay \textit{at al.}~\cite{mackay01a}.
They found that the choice of coin operator gave different
prefactors to the linear spreading rate of the quantum walk (compared to
quadratic classically) and showed some different symmetries for
different coin operators.

Here we present a more systematic (but by no means comprehensive)
investigation of the effects of different unbiased coin operators
combined with different initial states.  We consider mainly an unbounded,
regular, square lattice, but also consider the cases where
the edges are joined in either normal periodic boundary conditions
to give a torus, or twisted to give a Klein bottle.

\subsection{Quantum Walk on a Two Dimensional Lattice}

\begin{figure}
    \begin{minipage}{\columnwidth}
            \begin{center}
            \resizebox{\columnwidth}{!}{\includegraphics{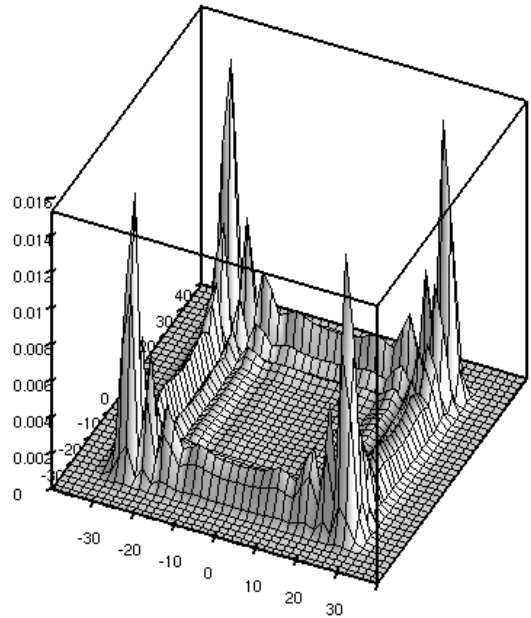}}
            \end{center}
            \caption{Distribution obtained after 40 steps of a quantum walk
		on a square lattice using a Hadamard coin operator
		and the symmetric initial state Eq.~(\ref{eq:sym0}).}
            \label{fig:had40-sym}
    \end{minipage}
\end{figure}

One obvious generalisation of a Hadamard coin to two spatial dimensions is
to take two Hadamard coins, one for left or right (\ket{L}, \ket{R}),
and one for up or down (\ket{U}, \ket{D}).
As shown in \cite{mackay01a}, this simply produces the same
pattern as the Hadamard coined walk on a line in both directions,
because the coin operator does not mix the two directions in any way,
see Fig.~\ref{fig:had40-sym}.
The standard deviation is the same for all choices of initial state
that produce a symmetric distribution, even maximally mixed, and is
$\sqrt{2}$ larger than the standard deviation for the walk on the
line, as noted by Mackay \textit{et al.}~\cite{mackay01a}.

\begin{figure}
    \begin{minipage}{\columnwidth}
            \begin{center}
            \resizebox{\columnwidth}{!}{\includegraphics{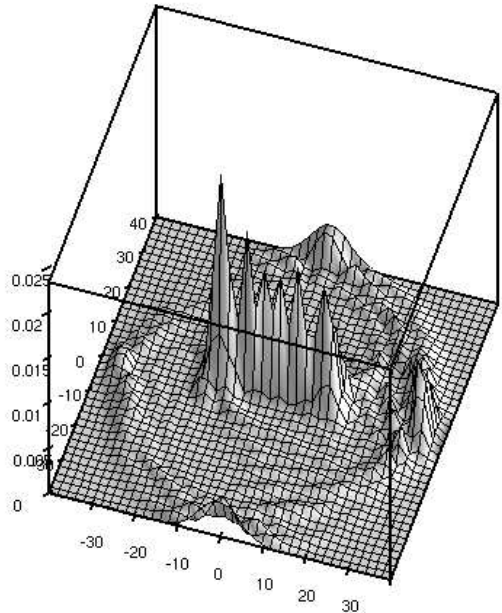}}
            \end{center}
            \caption{Distribution obtained after 40 steps of a quantum walk
		on a square lattice using a DFT coin operator
		and the symmetric initial state Eq.~(\ref{eq:sym0}).}
            \label{fig:dft40-typ}
    \end{minipage}
\end{figure}

\begin{figure}
    \begin{minipage}{\columnwidth}
            \begin{center}
            \resizebox{\columnwidth}{!}{\includegraphics{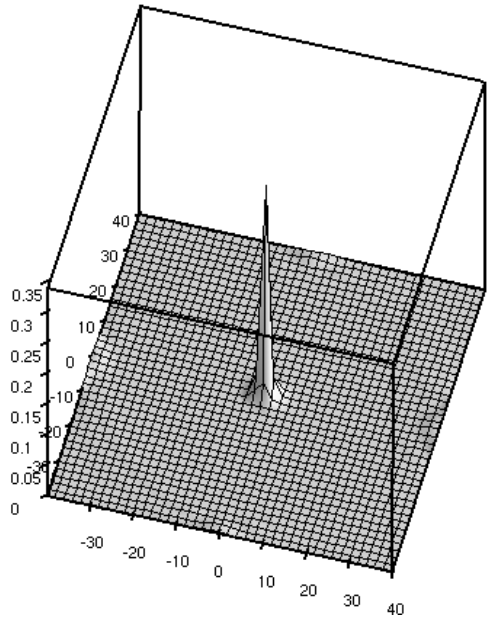}}
            \end{center}
            \caption{Distribution obtained after 40 steps of a quantum walk
		on a square lattice using a Grover coin operator
		and the symmetric initial state Eq.~(\ref{eq:sym0}).}
            \label{fig:grv40-typ}
    \end{minipage}
\end{figure}

More interesting are the degree-4 DFT coin,
\begin{equation}
\mathbf{C}_4^{(\text{D})}=\frac{1}{2}\left( \begin{array}{rrrr}
        1 & 1 & 1 & 1\\
        1 & i & -1 & -i\\
        1 & -1 & 1 & -1\\
        1 & -i & -1 & i\\
        \end{array} \right),
\label{eq:DFT4}
\end{equation}
and degree-4 Grover coin,
\begin{equation}
\mathbf{C}_4^{(\text{G})}=\frac{1}{2}\left( \begin{array}{rrrr}
        -1 & 1 & 1 & 1\\
        1 & -1 & 1 & 1\\
        1 & 1 & -1 & 1\\
        1 & 1 & 1 & -1\\
        \end{array} \right),
\label{eq:GROV4}
\end{equation}
(the only case where the Grover coin is unbiased),
both used in Ref.~\cite{mackay01a}.
Typical results for these coins and a symmetric initial coin state
\begin{eqnarray}
\ket{\psi_0^{(\text{sym})}} &=& \frac{1}{2}\left(\ket{LD}+i\ket{LU}+i\ket{RD}-\ket{RU}\right)\otimes\ket{0}\nonumber\\
&=& \frac{1}{2}\left(\ket{L}+i\ket{R}\right)\otimes\left(\ket{D}+i\ket{U}\right)\otimes\ket{0},
\label{eq:sym0}
\end{eqnarray}
with the particle starting at the origin
are shown in Figs. \ref{fig:dft40-typ} and \ref{fig:grv40-typ}.

As noted in Sec.~\ref{sec:2d},
there is essentially only one type of coin operator for the quantum
walk on a line, the Hadamard operator, with the full range of outcomes
accessible by adjusting the coin initial state.
For the walk on a two dimensional lattice, the situation is obviously
more complicated: the three coins illustrated so far give
distinctly different results whatever initial coin state is chosen.
The full range of possibilities is determined by the SU(4) group
structure of the unitary coin operator, but in order to sample the
possibilities numerically, we chose to look at unbiased coins (all elements
have modulus one half) and to further restrict those elements to
be $\pm 1/2$ or $\pm i/2$.  Choosing the leading diagonal entry to
be $+1/2$ leads to a set of 640 such unitary coin operators,
however, there is a high degree of redundancy if one groups all
results that are the same apart from rotation or reflection.
This can be done by using a simple initial state of (say)
$\ket{RU}$, and recording the second moment of the distribution.
The 640 coin operators then fall into just 10 types, with either
32, 64 or 128 coin operators of the original 640 in each type
(more symmetric distributions have fewer variations).
The Hadamard, Grover and DFT coin operators are all of different types.

We then varied the initial state of the coin, and looked for the maximum
and minimum second moments.  These always occured for symmetric
distributions (zero first moment), the second moment is thus
equal to the variance in these cases.
Our results contradict those of Mackay \textit{et al.}, whose choices
of initial states did not fully exploit the properties of the Grover coin.
Out of the ten types, the Grover type coin can produce both the
maximum and minimum possible second moments,
meaning that depending on the initial state,
it can either spread fastest or slowest from the starting point.
The distributions make clear why, see Fig.~\ref{fig:grv40-typ}.
They have an imperfect circular symmetry on the square lattice,
with a central spike, and a ring with something like the profile of
the distribution of the walk on a line superimposed on it.
The different initial states control how much of the distribution is
in the central spike and how much is in the ring, leading to the 
minimum and maximum values of the standard deviation.
In fact, most of the distribution ends up in the central spike,
except for exactly the right choice of initial state,
\begin{figure}
    \begin{minipage}{\columnwidth}
            \begin{center}
            \resizebox{\columnwidth}{!}{\includegraphics{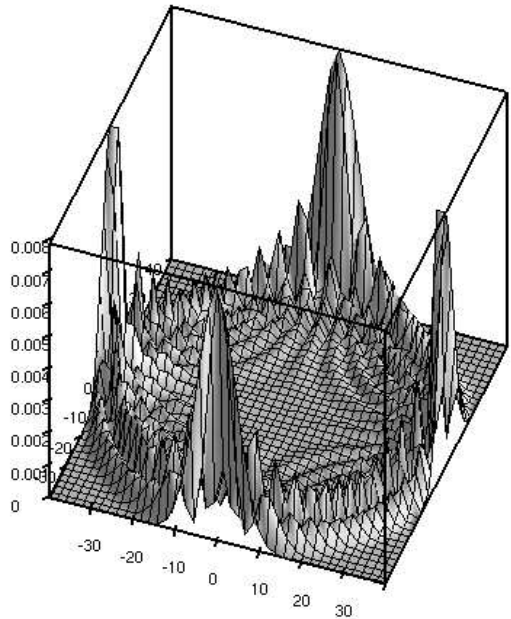}}
            \end{center}
            \caption{Distribution obtained after 40 steps of a quantum walk
		on a square lattice using a Grover coin operator
		and the symmetric initial state Eq.~(\ref{eq:symG}).}
            \label{fig:grv40-sym}
    \end{minipage}
\end{figure}
\begin{equation}
\ket{\psi_0^{(\text{G})}} = \frac{1}{2}\left(\ket{LD}-\ket{LU}-\ket{RD}+\ket{RU}\right).
\label{eq:symG}
\end{equation}
Figure \ref{fig:grv40-sym} shows the distribution produced from this
initial state, the contrast with Fig.~\ref{fig:grv40-typ} due to the
absense of the central spike is striking (though note the vertical
axes have different scales).
Shenvi \textit{et al.}~exploit this property of the Grover
operator in a different way in their quantum walk search algorithm
\cite{shenvi02a}.  Here they perturb the coin operator by applying a
different operation just at one marked vertex.  This causes an initially
uniform particle distribution over the whole lattice to converge on
the marked vertex, the reverse of a quantum walk starting at the origin
and spreading out.

\begin{figure}
    \begin{minipage}{\columnwidth}
            \begin{center}
            \resizebox{\columnwidth}{!}{\includegraphics{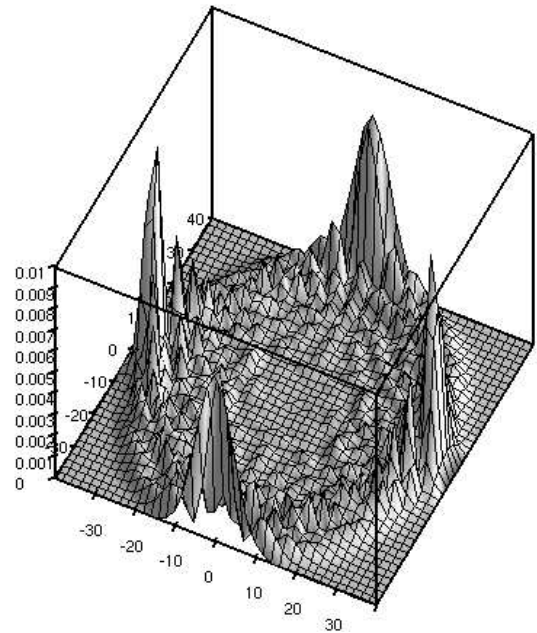}}
            \end{center}
            \caption{Distribution obtained after 40 steps of a quantum walk
		on a square lattice using a DFT coin operator
		and the symmetric initial state Eq.~(\ref{eq:symD}).}
            \label{fig:dft40-ring}
    \end{minipage}
\end{figure}

A DFT coin is not so symmetric (at most rotationally symmetric through
$\pi$, whereas both Hadamard and Grover coin operators can produce
distributions rotationally symmetric through $\pi/2$),
but with the right choice of initial condition, it too can produce
a ring shape with no central spikes, see Fig.~\ref{fig:dft40-ring}.
The initial state that produces this distribution is
\begin{equation}
\ket{\psi_0^{(\text{D})}} = \frac{1}{2}\left(\ket{LD}+\frac{1-i}{\sqrt{2}}\ket{LU}
	+\ket{RD}-\frac{1-i}{\sqrt{2}}\ket{RU}\right).
\label{eq:symD}
\end{equation}

The results presented by Mackay \textit{et al.}~\cite{mackay01a}
did not test a sufficiently wide range of initial states to draw
representative conclusions about the effects of entangled coins
on the distributions obtained for the quantum walks.
They attributed faster spreading to lack of entanglement between
the coin directions.  However, the differences we have found between
Hadamard coins and Grover or DFT coins are not in the degree of
spread per se, but in the extent to which this can be varied
simply through varying the initial coin state.
Both the Grover and DFT coins produce faster spreading than the
Hadamard coins with the initial states noted above.

\subsection{Cycles in two dimensions}
\label{ssec:d4wrap}

By joining a square or rectangular section of a two dimensional
lattice at opposite edges, the walk space becomes periodic
in both directions.  In one dimension there is only the
$N$-cycle, but in two dimensions the edges can be joined
directly, or twisted like a M\"{o}bius strip.  This gives
three structures that are two dimensional analogues of
the $N$-cycle, a torus, a closed M\"{o}bius strip, and a Klein bottle,
depending on whether none, one, or both pairs of the edges are joined twisted.
Periodic boundary conditions of these types are easy to implement
numerically.  We tested a range of such structures using the
same coins as for the walk on a lattice, and found similar
results to those for a walk on a $N$-cycle with respect
to mixing times and limiting distributions.
Further results for two dimensional cycles are presented at
the end of Sec.~\ref{sec:pd}.

\section{Graphs of higher degree}
\label{sec:hd}

For completeness, we mention that the hypercube, first studied
by Moore and Russell \cite{moore01a}, and later found by
Kempe \cite{kempe02a} to illustrate the possibility of an
exponential speed up with quantum walks, uses a higher dimensional
coin.  A hypercube with $2^N$ vertices has exactly $N$ connections to
each vertex and thus requires a $N$ dimensional coin Hilbert space.
Shenvi \textit{et al.}~\cite{shenvi02a} also based their
quantum walk search algorithm on a hypercube, though as they note,
other lattices, such as a square lattice, will do equally well.
The symmetry of the hypercube with a Grover coin
is such that with a symmetric initial state, the whole problem
may be mapped to a walk on a line with a variable coin operator, in the same
way as for the ``glued trees'' graph (see Sec.~\ref{sec:3d}).
Consquently, there is a wide range of possibilities with less symmetric
higher dimensional coins yet to be explored.

\section{Periodicity in Quantum Walks}
\label{sec:pd}

Systematic study of a quantum walk on a $N$-cycle,
(described in Sec.~\ref{ssec:2dcyc}), shows that among the smaller
values of $N$, a number of completely periodic walks arise.
This is the opposite property to mixing: here
the walk returns exactly to its initial state after a finite
number of steps $\Omega$, whereupon it repeats the same set of steps
and returns exactly again after $2\Omega$ steps, and so on.
There is no classical analogue of this property for random walks,
since in the classical case the dynamics are not deterministic.  
A classical random walk on a cycle will return to its starting state
at irregular, unpredictable times.
Note too, that this periodicity is not connected with whether the
limiting distribution is uniform or not, since here we are concerned with
exact return to the initial state, rather than the time-averaged
quantity in Eq.~(\ref{eq:limdef}).  Related ideas in continuous time
walks have been studied by Ahmadi \textit{et al.}~\cite{ahmadi02a},
where they are concerned with exact instantaneous uniform mixing,
rather than exact instantaneous return to the initial state.
\begin{figure}
    \begin{minipage}{\columnwidth}
            \begin{center}
            \resizebox{\columnwidth}{!}{\includegraphics{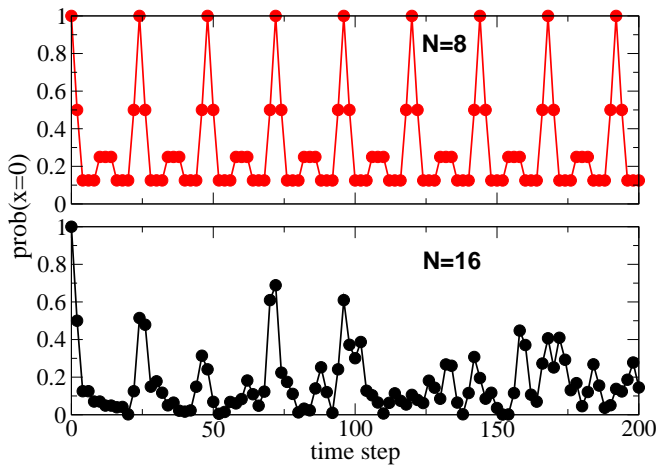}}
            \end{center}
            \caption{Probability of finding the particle at its
		initial position (x=0) for cycles of size $N=8$
		(upper) and $N=16$ (lower) plotted against the time
		step of the quantum walk using a Hadamard coin.
		Only even time steps are plotted, since for odd time
		steps the probability of finding the particle at an
		even numbered node is zero.}
            \label{fig:cycleperiod}
    \end{minipage}
\end{figure}

Using a Hadamard coin, the ``cycle'' of size $N=2$ is trivially periodic,
returning to its original state after two steps.  A cycle of size $N=4$ has
a period of eight steps.  This was first noted by
Travaglione and Milburn \cite{travaglione01a}.
The cycle with $N=8$ has a period of 24 steps, but $N=16$ is chaotic
and does not return to its initial state exactly even after many
thousands of steps.
This is illustrated in Fig.~\ref{fig:cycleperiod}, where the
probability of the particle being at its initial position is plotted
as a function of the time step.  A probability of one is an exact return
to the initial state (modulo the coin state, which is not shown here, but
does also, in fact, return to exactly the initial state).
If the coin is allowed to be biased, then a few more periodic examples
can be found, $N=6$ with period 12, and $N=10$ with period 60.
With judicious choice of phases in place of the Hadamard phases of
$\theta=\phi=0$,
$N=3$ has a period of 12, and $N=5$ has a period of 60, clearly
related to $N=6$ and $N=10$ respectively,
but these were the only odd-$N$ cycles we found.
These results are summarised in Table \ref{tab:period}.
\renewcommand{\arraystretch}{1.4}
\begin{table}
\caption{Known periods in a walk on a cycle. Coin phase $\delta=0$ unless specified.}
\label{tab:period}
\begin{tabular}{ccl}
$N$	&period $\Omega$& bias in coin $\rho$ \\
\hline
2	&	2	&	$\frac{1}{2}$ \\
3	&	12	&	$\frac{1}{3}$, $\delta=\frac{\pi}{3}$ \\
4	&	8	&	$\frac{1}{2}$ \\
5	&	60	&	$\left(\frac{\sin(\pi/6)}{\sin(\pi/5)}\right)^2$, $\delta=\frac{3\pi}{5}$ \\
6	&	12	&	$\frac{1}{3}$ \\
8	&	24	&	$\frac{1}{2}$ \\
10	&	60	&	$\left(\frac{\sin(\pi/6)}{\sin(\pi/5)}\right)^2 \simeq 0.7236$ \\
16	&\textit{chaotic}&	$\frac{1}{2}$ \\
\end{tabular}
\end{table}

The condition that must be satisfied for exact periodicity is
obtained from Eq.~(\ref{eq:linesol}), which also holds for the walk on
a cycle if the appropriate forms for the eigenvalues and eigenvectors are
substituted.  The wavefunctions $\ket{\psi_t}$
at two different times, $t$ and $t+\Omega$ are set equal, giving
\begin{equation}
(\lambda_k^{\pm})^{\Omega} = 1 ~~~\forall~k \in \{0,1\dots N-1\}.
\end{equation}
Using Eq.~(\ref{eq:lambdakcycle}) gives
\begin{eqnarray}
(\delta + \omega_k)\Omega &=& 2\pi j_+, \nonumber\\
(\delta - \omega_k + \pi)\Omega &=& 2\pi j_-,
\end{eqnarray}
where $j_{\pm}$ are integers.
Substituting these into Eq.~(\ref{eq:omegakcycle}) gives
\begin{equation}
\cos\left(\frac{\pi j}{\Omega}\right) = 
\sqrt{\rho}\cos\left(\frac{2\pi k}{N} - \frac{\pi m}{\Omega}\right)
~~~\forall~k,
\label{eq:period}
\end{equation}
where $\rho$ is the bias in the coin operator, $m$ is an integer specifying
the relative phases in the coin operator through
$m\pi/\Omega-\pi/2 = \delta=(\theta+\phi)/2$,
$k$ is the integer Fourier variable, and $j$ is an integer that
can be different for each $k$, but must be odd or even to match whether $m$
is odd or even.
Clearly, the larger $N$ is, the harder it is to find solutions for
Eq.~(\ref{eq:period}) for all $k$ at the same time (apart from the
trivial solutions for $\rho=0$ or 1).  We do not know if we have
found all possible solutions that give periodic quantum walks on a cycle,
but we conjecture that there are only a finite number of such solutions
and that we have found nearly all if not all of them.

We also studied periodicity on two dimensional cycles, as described
in Sec.~\ref{ssec:d4wrap}.
With a Hadamard coin and a torus made from suitable
small dimensions, periodicity is also obtained
in the cases predictable from Table \ref{tab:period}.
For a closed M\"{o}bius strip or Klein bottle, the twisted dimension
is only periodic if the size is half that in Table \ref{tab:period},
because the twist causes the walk to traverse the cycle twice
before returning to its initial state.  The Grover coin shows
the same periodicities as the Hadamard coin.  However, a DFT
coin only shows perodicity for a torus of dimensions $4\times4$,
and not at all on the twisted surfaces.  This is due to the
asymmetry of the DFT coin compared to the Grover and Hadamard coins.
During the double circuit of the twisted surface, the wavefunction
interfers with a mirror image of itself, so periodicity will only
be observed with coins that produce suitably mirror symmetric
distributions.

\section{Summary}
\label{sec:Conc}

We have studied discrete, coined quantum walks on regular lattices,
in one and two spatial dimensions (graphs with vertices of degree
two, three and four).  Both the bias (away from equal probability
of choosing each direction) and the phases in the coin operator, and
the initial state of the coin, can be used to control the evolution
of the quantum walk.  In a quantum walk on a line, we have found
a biased initial state of the coin which produces a higher degree of
asymmetry than the simple $\ket{L}$ or $\ket{R}$ initial states.
The same bias produces a symmetric distribution when combined with
the opposite phase between the coin components.  This illustrates
two distinct ways to obtain the same symmetric distribution, by
interference, and by combination of two orthogonal biased distributions
each a mirror image of the other.
In a quantum walk on a cycle, we have determined the condition for
mixing to a uniform limiting distribution, for a general coin operator
and initial state.  Non-uniform limiting distributions are highly sensitive
to decoherence, so to make use of the properties of such walks, it will be
best to measure effects that occur after a reasonably short number
of steps of the walk.
Quantum walks of degree three have a more interesting choice of coin operators,
an example in which a Grover coin solves a problem (``glued trees'')
efficiently, while a DFT coin stays nearer the starting point than
even a classical random walk illustrate the range of possibilities
to be explored.
Numerical study of regular lattices in two dimensions (degree-4 graphs)
show that the Grover and DFT coins have interesting properties independent of 
the symmetry of the lattice (circular spreading on a square lattice).
Suitable choice of initial state makes the Grover coin spread fastest
or slowest out of all the coin operators tested, in contrast to
the conclusions in \cite{mackay01a}, where only a few initial states were
tested.
Finally, a small set of exactly periodic quantum walks on cycles of
sizes 2, 3, 4, 5, 6, 8, and 10 have been found, and the condition on which this
exact periodicity depends derived.  Such periodicities are of interest
in their own right, and we also suggest that it may be possible to exploit
them to pick out small scale regularities in larger structures.

\begin{acknowledgments}
We thank Julia Kempe, Peter Knight, and Jiannis Pachos for useful discussions.
VK thanks the Mathematical Science Research Institute in Berkeley, CA, for
hospitality during the Quantum Computing Program in Fall 2002, and the
participants of that programme for many stimulating discussions.
BT and VK funded by the UK Engineering and Physical Sciences
Research Council.
\end{acknowledgments}


\bibliography{qrw}


\end{document}